\documentclass[sigconf]{acmart}
\AtBeginDocument{%
  \providecommand\BibTeX{{%
    \normalfont B\kern-0.5em{\scshape i\kern-0.25em b}\kern-0.8em\TeX}}}

\newcommand{\eg}{\emph{e.g., }}

\newcommand{\sth}[1]{\textcolor{black}{#1}}
\newcommand{\sthn}[1]{\textcolor{black}{#1}}
\newcommand{\xzj}[1]{\textcolor{black}{#1}}
\usepackage{enumitem}
\usepackage{verbatim}
\usepackage[ruled]{algorithm2e}
\usepackage{multirow}
\usepackage{subfigure} 
\usepackage{ragged2e}
\usepackage{caption}
\usepackage{subcaption}
\usepackage{graphicx}
\usepackage[normalem]{ulem}
\useunder{\uline}{\ul}{}
\usepackage{amsmath}
\usepackage{setspace}
\usepackage{bm}

\copyrightyear{2024}
\acmYear{2024}
\setcopyright{acmlicensed}\acmConference[CIKM '24]{Proceedings of the 33rd ACM International Conference on Information and Knowledge Management}{October 21--25, 2024}{Boise, ID, USA}
\acmBooktitle{Proceedings of the 33rd ACM International Conference on Information and Knowledge Management (CIKM '24), October 21--25, 2024, Boise, ID, USA}
\acmDOI{10.1145/3627673.3679922}
\acmISBN{979-8-4007-0436-9/24/10}

\begin{document}



\title{Preliminary Study on Incremental Learning for Large Language Model-based Recommender Systems}

\author{Tianhao Shi}
\affiliation{%
  \institution{University of Science and Technology of China}
    \city{Hefei}
  \country{China}}
\email{sth@mail.ustc.edu.cn}

\author{Yang Zhang}
\authornote{Corresponding authors.}
\affiliation{%
  \institution{National University of Singapore}
    \city{Singapore}
  \country{Singapore}}
  \email{zyang1580@gmail.com}

\author{Zhijian Xu}
\affiliation{%
  \institution{University of Science and Technology of China}
    \city{Hefei}
  \country{China}}
\email{zjane@mail.ustc.edu.cn}

\author{Chong Chen}
\affiliation{%
  \institution{Huawei Cloud BU}
  \city{Shenzhen}
  \country{China}}
\email{chenchong55@huawei.com}

\author{Fuli Feng}
\affiliation{%
  \institution{University of Science and Technology of China}
    \city{Hefei}
  \country{China}}
\email{fulifeng93@gmail.com}

\author{Xiangnan He}
\affiliation{%
  \institution{University of Science and Technology of China}
    \city{Hefei}
  \country{China}}
\email{xiangnanhe@gmail.com}

\author{Qi Tian}
\affiliation{%
  \institution{Huawei Cloud BU}
  \city{Shenzhen}
  \country{China}}
\email{tian.qi1@huawei.com}






\renewcommand{\shortauthors}{Tianhao Shi, et al.}

\begin{abstract}
Adapting Large Language Models for Recommendation (LLM4Rec) has shown promising results. However, the challenges of deploying LLM4Rec in real-world scenarios remain largely unexplored. In particular, recommender models need incremental adaptation to evolving user preferences, while the suitability of traditional incremental learning methods within LLM4Rec remains ambiguous due to the unique characteristics of Large Language Models (LLMs).

    
In this study, we empirically evaluate two commonly employed incremental learning strategies (full retraining and fine-tuning) for LLM4Rec. Surprisingly, neither approach shows significant improvements in the performance of LLM4Rec. Instead of dismissing the role of incremental learning, we attribute the lack of anticipated performance enhancement to a mismatch between the LLM4Rec architecture and incremental learning: LLM4Rec employs a single adaptation module for learning recommendations, limiting its ability to simultaneously capture long-term and short-term user preferences in the incremental learning context. To test this speculation, we introduce a Long- and Short-term Adaptation-aware Tuning (LSAT) framework for incremental learning in LLM4Rec. Unlike the single adaptation module approach, LSAT utilizes two distinct adaptation modules to independently learn long-term and short-term user preferences. Empirical results verify that LSAT enhances performance, thereby validating our speculation. 
We release our
code at: \url{https://github.com/TianhaoShi2001/LSAT}.
\end{abstract}
\vspace{-2pt}

\ccsdesc[500]{Information systems~Recommender systems}

\vspace{-2pt}
\keywords{Large Language Models, Model Retraining, Incremental Learning}


\maketitle

\section{Introduction}

The emergence of Large Language Models~\cite{surveyLLM}, equipped with extraordinary capabilities in text comprehension and generation, \sthn{has been successfully applied in various domains like Robotics~\cite{palm,singh2023progprompt} and Computer Vision~\cite{cvllm, berrios2023towards}.} Inspired by this success, 
there is growing interest in using LLMs for recommendations in both academia~\cite{LLMREC1,llmrec2,llmrec3} and industry~\cite{lin2023can}.
\sthn{Among current efforts, tuning LLMs with recommendation-specific data
using LoRA~\cite{lora} (a well-known efficient fine-tuning method)
has yielded promising results~\cite{tallrec,collm,bigrec,rella,zhang2024text}, underscoring the substantial potential of LLM4Rec in real-world applications.}
However, the challenges associated with the practical deployment of LLM4Rec remain explored, particularly considering the unique characteristics of LLMs.
When deploying a recommender system in real-world scenarios, one of the primary challenges is ensuring the recommender models can adapt incrementally to evolving user preferences and environments~\cite{sml,period-update,incrementalSurvey}.
\sth{This adaptation is crucial as user feedback streamingly flows in, requiring the recommender model to be incrementally updated with the latest data to achieve timely personalization.
For traditional recommendation models, the critical role of incremental learning and associated challenges have been extensively researched~\cite{sml,incrementalSurvey,period-update}.
However, when it comes to LLM4Rec, the issues related to incremental learning lack adequate attention. 
The unique characteristics of LLM4Rec, such as its 
massive parameters
and its high tuning cost~\cite{llmrec-benchmark},
may introduce new challenges or insights that require thorough examination.}
\sth{In this study, we first empirically examine how incremental learning impacts the performance of LLM4Rec.}
Considering the broad adoption of LoRA
in developing LLM4Rec models~\cite{tallrec,bigrec,rella,collm}
, and acknowledging LoRA's efficiency and effectiveness~\cite{liu-peft}, our research focuses on this specific type of LLM4Rec. 
We examine two commonly used incremental learning strategies: 1) full retraining~\cite{period-update}, which involves periodic retraining using complete historical data and new data, and 2) fine-tuning~\cite{rendle2008online,wang2018neural}, which updates the model based solely on new data. 
\sth{Based on our empirical results, we find that both full retraining and fine-tuning have a minimal impact on the performance of LLM4Rec.}
\sth{These results emphasize that LLM4Rec exhibits good generalization capabilities even under delayed updates; however, they also suggest that incremental learning might not lead to performance improvements for LLM4Rec.}
Based on our empirical results, executing incremental learning appears to be unnecessary for LLM4Rec.
This is somewhat surprising, as user preferences do change over time, and a recommender system should adapt to these changes~\cite{sml,period-update,longshort}. We speculate that the lack of anticipated performance improvements may be attributed to a mismatch between the LoRA architecture and incremental learning: LoRA avoids training the entire model and instead tunes a low-rank adaptation module~\cite{liu-peft} with recommendation data, while a single LoRA module may have the inability to simultaneously emphasize long-term and short-term user \xzj{preferences} under incremental learning.
\sth{
Specifically, for full retraining, the LoRA module might emphasize long-term preferences but overlook short-term ones, given the substantial volume of historical data compared to the new data~\cite{jang2021towards}.
For fine-tuning, the LoRA module may forget previous knowledge due to catastrophic forgetting~\cite{luo2023empirical}, leading to a decline in performance.
}

To test our speculation, we develop a modified updating method called \textit{Long- and Short-term Adaptation-aware Tuning} (LSAT). 
This method utilizes two LoRA modules to separately learn long-term and short-term user \xzj{preferences} and then integrates them to merge the different types of \xzj{preferences}. 
\sthn{During each update, the short-term LoRA module is
temporarily retrained using solely new data to focus on the latest evolving preferences. In contrast, given the robust generalization capabilities exhibited by the long-term LoRA even with delayed updates, it remains fixed once it has been sufficiently trained or retrained at a relatively gradual frequency to conserve training costs. We conduct a comparison between LSAT, full retraining, and fine-tuning methods. Extensive results demonstrate that LSAT brings performance enhancements, confirming the validity of our speculation. 
Nevertheless, at present, LSAT only explores incremental learning from the perspective of LoRA capacity.
To comprehensively understand and address the issue, further investigation in various directions is still necessary.}

The main contributions are summarized as follows:
\vspace{-2pt}
\begin{itemize}[leftmargin=*]
    \item New Problem: This work marks the inaugural investigation into incremental learning for LLM4Rec, furnishing practical insights for the real-world deployment of LLM4Rec.

    \item New Finding: Our empirical results underscore that the common incremental learning methods (full retraining and fine-tuning) do not clearly enhance the performance of LoRA-based LLM4Rec.

    \item Proposal: We propose that using separate LoRA modules to capture long-term and short-term preferences can enhance the performance of LLM4Rec in incremental learning, offering valuable insights from the perspective of the capacity of the LoRA module.
\end{itemize}
\section{RELATED WORKS}
\noindent \textbf{LLM-based recommender.} 
Adapting LLMs as a recommender has gained substantial attention.
In-context learning enables LLMs to provide recommendations without explicit training~\cite{wang2023zero,chen2023palr}. 
Nevertheless, due to the lack of recommendation-specific knowledge during pre-training, applying instruction tuning~\cite{instruction-tuning} with recommendation data helps LLMs achieve much better recommendation performance~\cite{tallrec,bigrec}.
Among tuning methods, InstructRec~\cite{zhang2023recommendation} fine-tunes all LLM's parameters, while most approaches employ parameter-efficient fine-tuning (PEFT)~\cite{liu-peft} to avoid adjusting the extensive parameters of LLMs~\cite{tallrec,bigrec,collm,rella}. Within PEFT, LoRA is adopted by the majority of LLM4Rec models~\cite{tallrec,bigrec,collm,rella} due to its good convergence and accuracy~\cite{liu-peft}. Considering the broad adoption of LoRA in developing LLM4Rec models, this work explores 
the issues of incremental learning in LoRA-based LLM4Rec.

\vspace{2pt}
\noindent \textbf{Incremental learning in recommendation.}
Incremental learning is crucial for recommender models due to new users/items and changing user preferences\cite{sima2022ekko,period-update}.
Representative methods for incremental updates include 1) full retraining~\cite{period-update}, which retrains models with both new and historical data, achieving high accuracy with extensive training cost; 
2) fine-tuning~\cite{rendle2008online,wang2018neural}, which updates models exclusively with the latest data, offering efficiency but facing potential forgetting; 
3) sample-based methods~\cite{diaz2012real,wang2018streaming}, which update models with new data and a sampled subset of historical data, where the sampled data is expected to retain long-term preference signals; 
and 4) meta-learning-based methods~\cite{sml,longshort}, utilizing meta-learning to optimize the model for better future serving.
 Unlike prior works focusing on traditional recommender models,  this work explores incremental learning within LLM4Rec.
\section{Preliminaries}

In this study, we explore incremental learning in LLM4Rec. 
Our investigation is centered around a representative LLM4Rec model known as TALLRec~\cite{tallrec}, chosen due to the widespread adoption of its tuning paradigm~\cite{bigrec,collm}. Next, we briefly introduce TALLRec and incremental learning in the recommendation.

\noindent $\bullet$ \textbf{TALLRec.} 
To align LLMs with recommendations,
TALLRec utilizes instruction tuning~\cite{instruction-tuning}. This involves organizing historical interaction data into textual instructions and responses and then fine-tuning LLMs using this structured data to improve recommendation performance.
Notably, TALLRec adopts LoRA~\cite{lora} for efficient tuning, which freezes pre-trained LLM parameters and integrates lightweight trainable matrices.
Specifically, LoRA introduces a pair of rank-decomposed weight matrices to each pre-trained weight matrix $W \in \mathcal{R}^{d\times k}$, formally $W+AB$, where $A \in \mathcal{R}^{d\times r}$ and $B \in \mathcal{R}^{r\times k}$ are added learnable LoRA matrices ($r\ll \min(d,k)$).
%

\begin{figure}[t]
    \centering
    \includegraphics[width=0.45\textwidth]{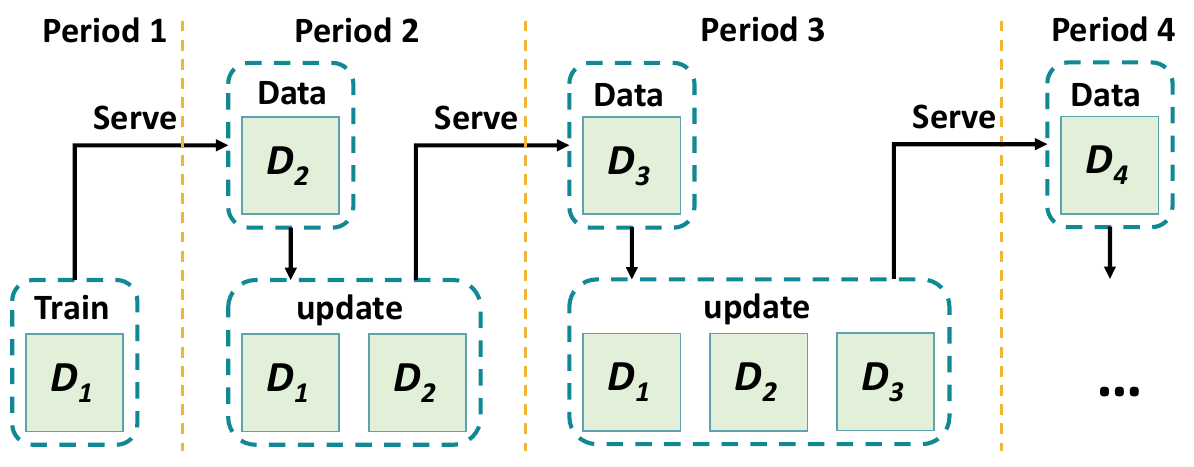}
    \vspace{-13pt}
    \caption{Incremental learning process in recommendation.}
    \vspace{-10pt}
    \label{fig:periodic-update}
\end{figure}

\noindent $\bullet$ \textbf{Incremental learning in recommendation.} 
To signify the incremental learning process, following~\cite{sml,period-update}, we represent the data stream as $\{\mathcal{D}_1, \mathcal{D}_2, \dots, \mathcal{D}_t, \dots \}$, where $\mathcal{D}_{t}$ represents the collected data at time period $t$. The length of a period may vary (\eg daily, weekly) based on system requirements.
At each period $t$, the model has access to new data $\mathcal{D}_{t}$ and all previous data for updating, and the updated model needs to serve for the near future $\mathcal{D}_{t+1}$.
This paper uses two representative incremental learning strategies: full retraining~\cite{period-update}, updating the model with both new data $\mathcal{D}_t$ and entire historical data \{$\mathcal{D}_{1}, \mathcal{D}_{2}, \dots, \mathcal{D}_{t}\}$; and fine-tuning~\cite{rendle2008online,wang2018neural}, utilizing only the latest data $\mathcal{D}_t$ for the update.
Figure~\ref{fig:periodic-update} illustrates the process of incremental learning in recommender systems.


\section{Empirical Explorations}
In this section, we conduct experiments to explore the impact of the commonly employed incremental learning methods on TALLRec.

\subsection{Experiemental Settings}
\label{sec:setting1}
\noindent\textbf{Datasets.} We conduct experiments on two representative datasets: MovieLens-1M (ML-1M)~\cite{movielens}, which is a movie rating dataset collected by GroupLens Research, and Amazon-Book~\cite{amazonbook}, which includes user reviews of books in Amazon. 
In ML-1M and Amazon-Book, ratings range from 1 to 5. 
Following~\cite{tallrec,collm}, interactions with ratings $\geq$ 4 are positive; others are negative.  
To pre-process the data, we adopt the approach from TALLRec~\cite{tallrec}, converting ratings to binary labels and excluding users with less than 10 interactions.
To assess incremental learning's impact, we divided the data chronologically based on interaction timestamps. For ML-1M, we use data from Dec. 2000 to Feb. 2003, creating 20 periods of 10,000 samples. 
Similarly, for Amazon-Book, we keep the data from Mar. 2014 to May. 2018, and sampled 20\% of users, resulting in 328,168 samples, which were then divided into 20 two-month periods. 
For each period, we train models on the initial 90\% of the data and validate them on the remaining 10\%.
Table~\ref{tab:statistics1} presents the dataset statistics.



\noindent \textbf{Models.}
Due to the diversity of incremental learning algorithms, we select the two most representative incremental learning strategies, full retraining, and fine-tuning for updating TALLRec.
We also evaluate how these two update methods affect five traditional recommendation models: 1) MF~\cite{koren2009matrix}, which is a latent factor-based collaborative filtering, 2)DeepMusic~\cite{deepmusic}, which is a content-based recommendation model, 3) GRU4Rec~\cite{gru4rec}, which is an RNN-based sequential recommender, 4) Caser~\cite{tang2018personalized}, which uses CNN to model sequence patterns, and 5) SASRec~\cite{kang2018self}, which employs a self-attention mechanism to grasp sequential patterns.
Additionally, we evaluate BookGPT~\cite{bookgpt}, an in-context learning-based method, which is tuning-free and always incorporates the latest data into prompts. 

\begin{table}[t] 
\centering
\caption{Statistics of the evaluation datasets.}
\label{tab:statistics1}
\vspace{-10pt}
\setlength{\tabcolsep}{1.2mm}{
\resizebox{0.42\textwidth}{!}{
\begin{tabular}{ccccc}
\hline
Dataset & \# Users & \# Items & \# Instances & Density \\ \hline
ML-1M   & 1,813     & 3,503     & 200,000       & 3.1491\%   \\ \hline
Amazon-Book  & 28,427    & 12,680   & 328,168       & 0.0869\%   \\ \hline
\end{tabular}}}
\vspace{-0.2cm}
\end{table}

\begin{figure}[t]
    \centering
    \subfigure
{\label{fig:a}\includegraphics[width=0.225\textwidth]{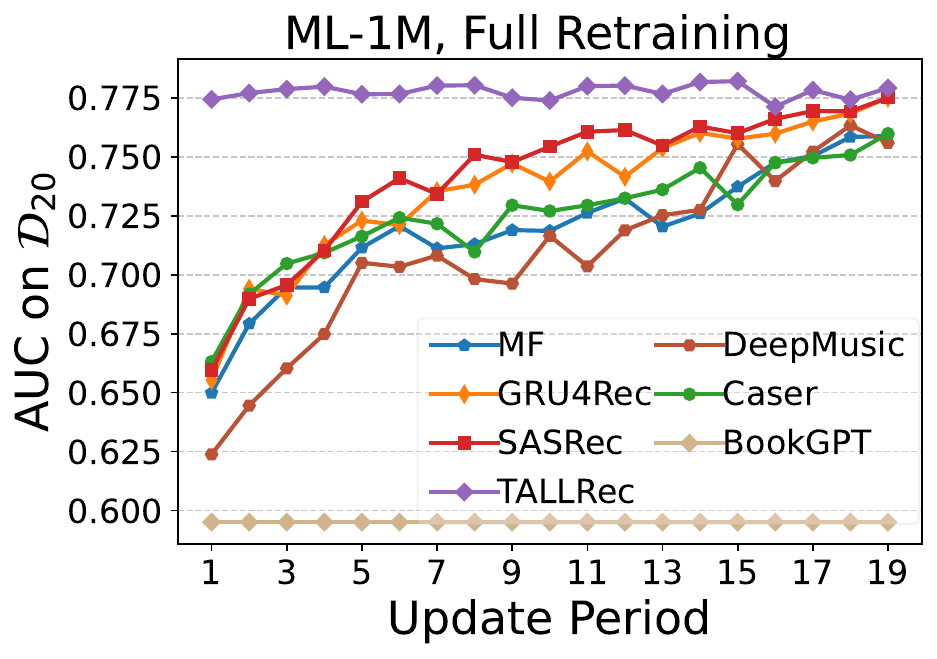}}
    \subfigure
    {\label{fig:b} \includegraphics[width=0.225\textwidth]{
    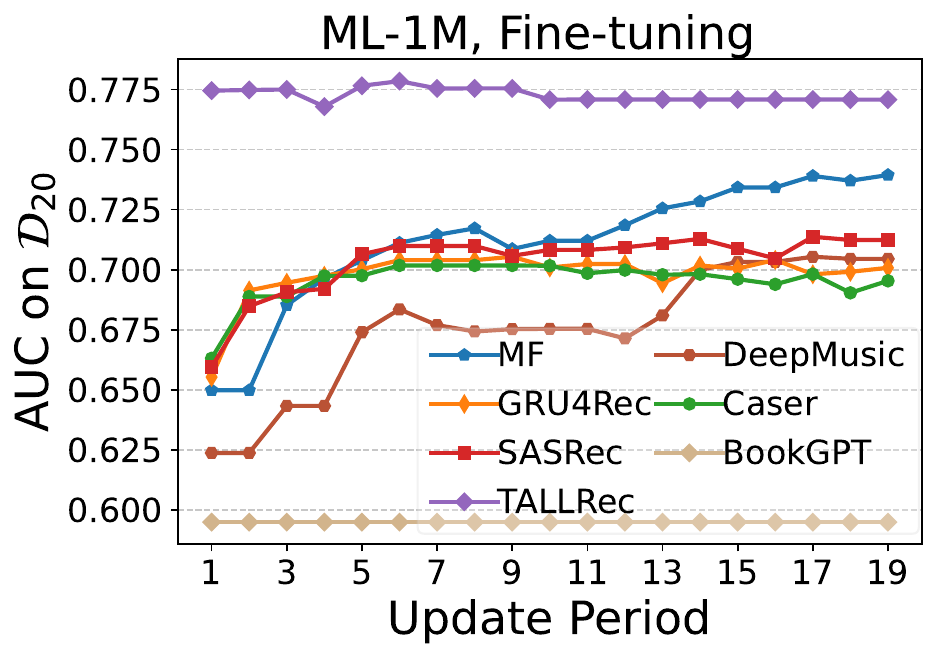}}

        \vspace{-6pt}
        \subfigure
        {\label{fig:b} \includegraphics[width=0.225\textwidth]{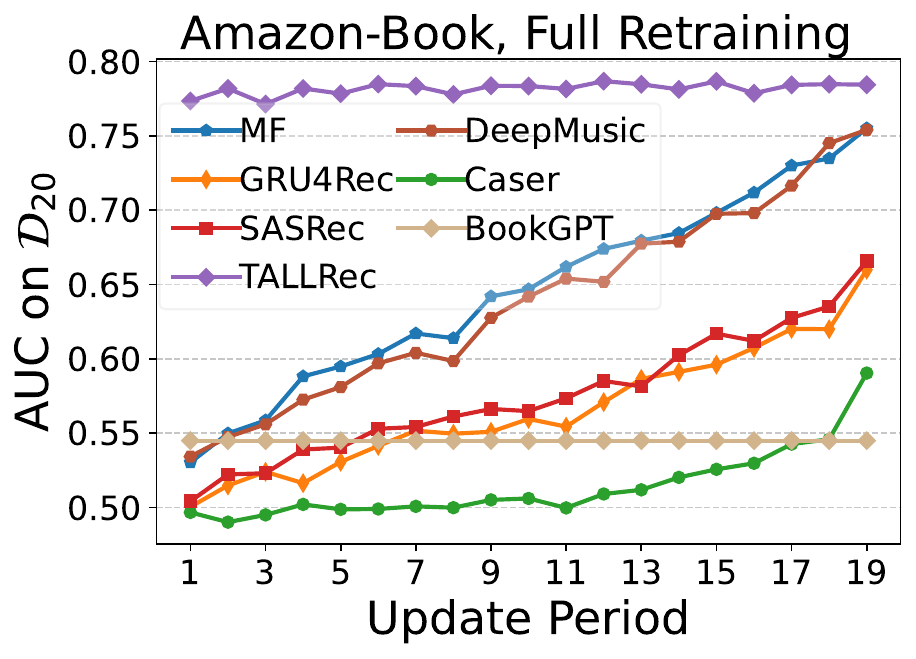}}
            \subfigure
            {\label{fig:b} \includegraphics[width=0.225\textwidth]{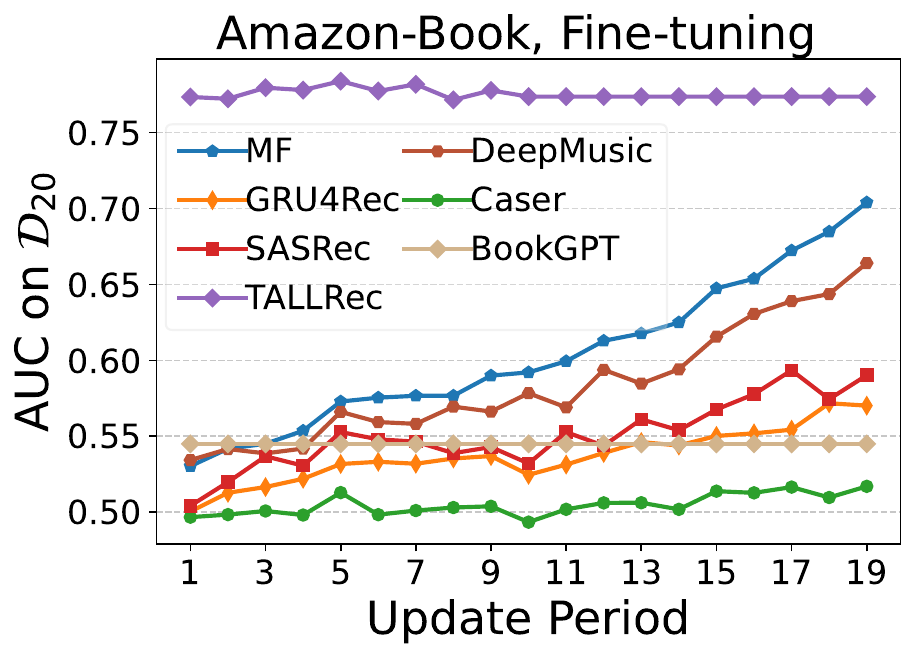}}
            \vspace{-10pt}
    \caption{Performance of TALLRec, BookGPT and traditional models obtained at different update periods on $\mathcal{D}_{20}$.}
    \vspace{-10pt}
    \label{fig:preliminary-exp}
\end{figure}

\noindent \textbf{Evaluation metrics and hyper-parameters.}
Following TALLRec~\cite{tallrec}, we use AUC~\cite{auc} to evaluate recommendation performance.
TALLRec is deployed based on LLaMA-7B~\cite{touvron2023llama}, and BookGPT on GPT-3.5-turbo, with settings aligned with the original papers.
We optimize all traditional models using the Adam optimizer~\cite{kingma2014adam} with the MSE loss, employing a learning rate of 1e-3, batch size of 256, embedding size of 64, and weight decay of 1e-5 (tuned results). 

\subsection{Incremental Learning's Impact on LLM4Rec}

\noindent \textbf{Overall results.}
To evaluate the effect of incremental learning, the model undergoes continuous updates until the 19th period. Then, we evaluate the performance of the model obtained at each update period on the data  $\mathcal{D}_{20}$ of the 20th period, and plot the performance curve against the update periods in Figure~\ref{fig:preliminary-exp}, where we can find: 
\begin{itemize}[leftmargin=*]
\vspace{-2pt}
\item For traditional models,
timely updates could enhance their performance, particularly with full retraining. 
Fine-tuning is usually less effective than full retraining, and its effectiveness may diminish over time, possibly due to the forgetting issue.
\item TALLRec significantly outperforms BookGPT, suggesting that while in-context learning can utilize the latest data without additional training, its performance is limited. 
\item Unlike traditional models, the performance of TALLRec remains relatively unaffected by both full retraining and fine-tuning.
This underscores its ability to excel in generalizing to new data.
However, it also indicates that timely incremental learning does not enhance LLM4Rec's performance.
\vspace{-2pt}
\end{itemize}

Our initial findings suggest that incremental learning has a limited impact on the performance of LLM4Rec.
Subsequently, we conduct a more detailed analysis to explore the effects of incremental learning on LLM4Rec, focusing on two key aspects as highlighted in~\cite{period-update}:
1) timely incorporation of new items and users, and 2) adaptation to changing user preferences.

\begin{figure}[t]
    \centering
    \subfigure
{\label{fig:a}\includegraphics[width=0.225\textwidth]{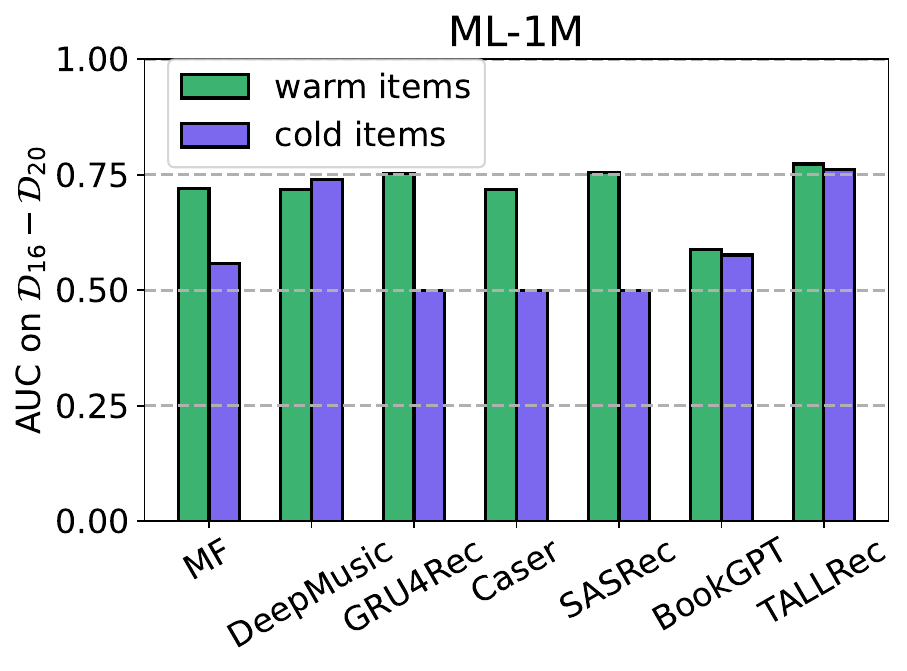}}
    \subfigure
    {\label{fig:b} \includegraphics[width=0.225\textwidth]{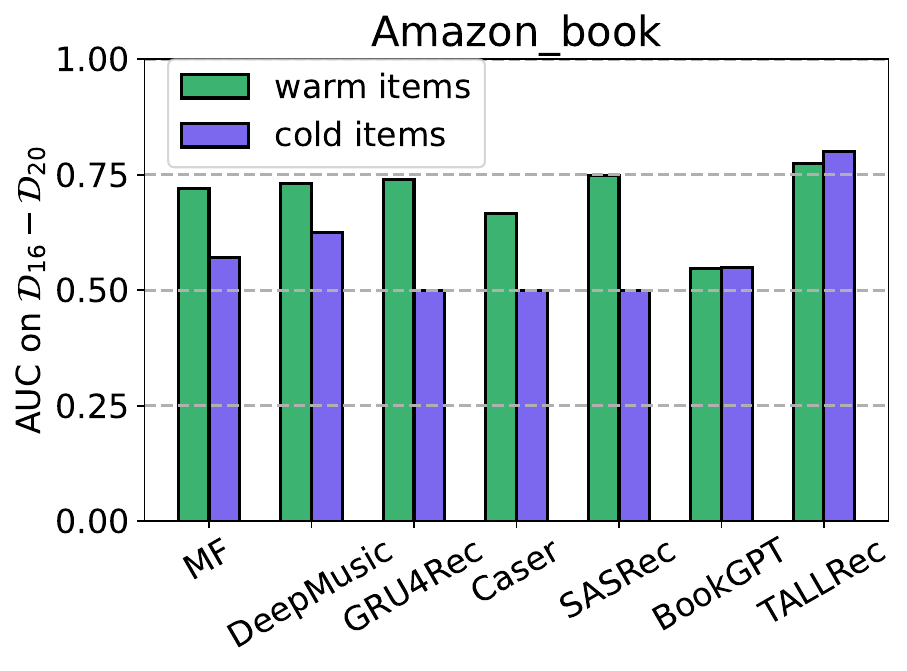}
    }
    \vspace{-10pt}
    \caption{Performance comparison between TALLRec and baselines on warm items and cold items. All models are trained on $\mathcal{D}_{1}-\mathcal{D}_{15}$ and tested on $\mathcal{D}_{16}-\mathcal{D}_{20}$.
    }
    \vspace{-9pt}
    \label{fig:cold}
\end{figure}

\noindent\textbf{Cold-start items evaluation.}
In examining the first aspect, we compare the performance of TALLRec and traditional models on warm and cold items (trained on $\mathcal{D}_{1}$--$\mathcal{D}_{15}$, tested on $\mathcal{D}_{16}$--$\mathcal{D}_{20}$) in Figure~\ref{fig:cold}. 
 From the figure, we can find collaborative filtering methods exhibit near-random guessing (AUC=0.5) for cold items, indicating they would suffer performance deterioration without updates due to an increased number of cold items~\cite{period-update}.
In contrast, TALLRec's proficiency in general language understanding enables accurate recommendations for cold items. 
Hence, from the perspective of cold items, incremental learning has a smaller impact on LLM4Recs compared to traditional models.

\begin{figure}[t]
    \centering
    \subfigure
{\label{fig:a}\includegraphics[width=0.225\textwidth]{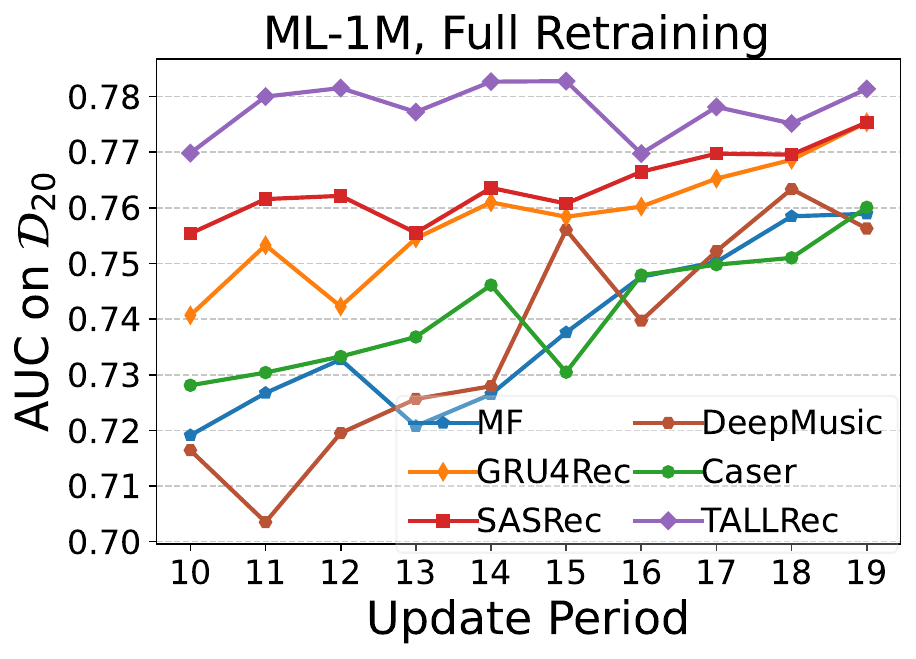}}
    \subfigure
    {\label{fig:b} \includegraphics[width=0.225\textwidth]{
    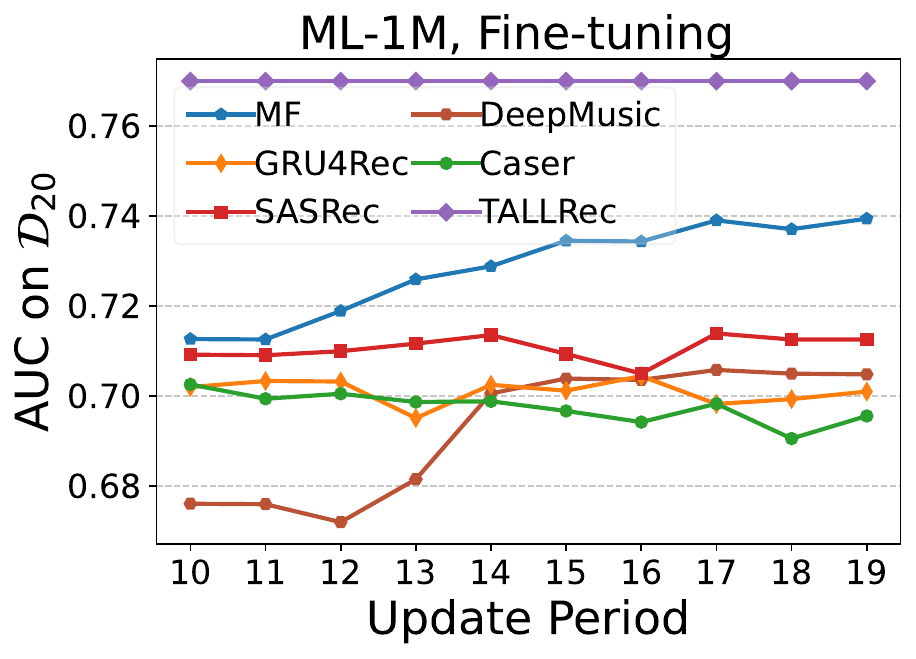}}

        \vspace{-6pt}
        \subfigure
        {\label{fig:b} \includegraphics[width=0.225\textwidth]{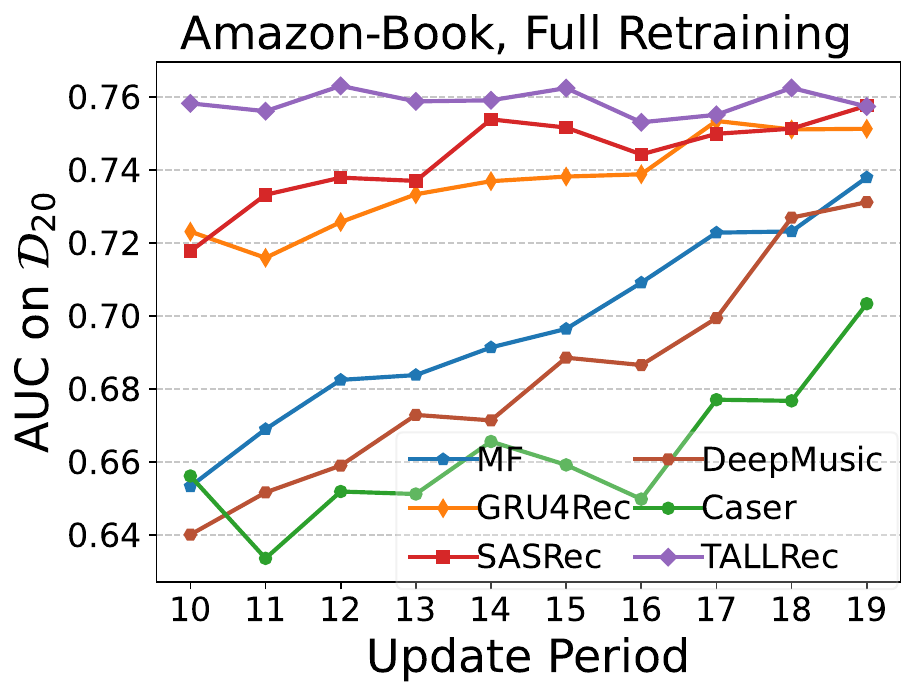}}
            \subfigure
            {\label{fig:b} \includegraphics[width=0.225\textwidth]{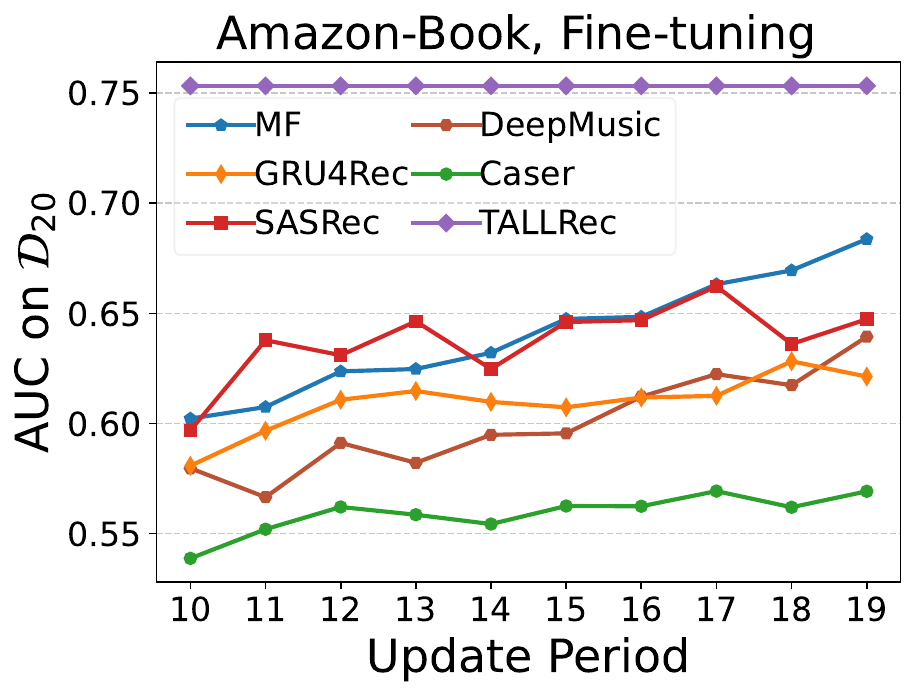}}
            \vspace{-10pt}
    \caption{Performance of TALLRec and traditional models obtained at different update periods for warm items on $\mathcal{D}_{20}$.
    }
    \vspace{-10pt}
    \label{fig:heatmap}
\end{figure}

\noindent\textbf{Dynamic preference on warm items.} 
We then explore whether incremental learning improves the performance of LLM4Rec by adapting to the latest user preferences.
Toward this, we filter cold items 
(items do not appear in $\mathcal{D}_{1} $--$\mathcal{D}_{10}$ but appear in $\mathcal{D}_{20}$)
and evaluate the performance of warm items on $\mathcal{D}_{20}$ under different periods in Figure~\ref{fig:heatmap}.
We find that incremental learning always improves the performance of traditional models on warm items, except for fine-tuning on ML-1M due to forgetting issues.   
However, both full retraining and fine-tuning cannot enhance TALLRec's performance on warm items.
We suggest this could be due to the inability of a single LoRA to capture both long-term and short-term user preferences simultaneously. Full retraining may focus more on long-term preferences due to the larger quantity of historical data~\cite{jang2021towards}. Fine-tuning might prioritize short-term preferences in new data while forgetting previous knowledge~\cite{luo2023empirical}. Consequently, both full-retraining and fine-tuning fail to achieve performance improvements by adapting to the latest preferences.

\section{LSAT}
\label{sec:LSAT}
We have observed that both full retraining and fine-tuning do not effectively improve LLM4Rec's performance.
We posit that a single LoRA module may struggle to simultaneously capture both long-term and short-term user preferences.
This insight draws inspiration from advancements in leveraging multiple LoRA modules to handle distinct tasks or domain knowledge~\cite{chronopoulou2023adaptersoup,zhang2023composing}.
Considering the divergence between long-term and short-term user preferences, it may be necessary to employ separate LoRA modules to capture them individually. To validate this speculation, we develop a new method called LSAT, employing two dedicated LoRA modules --- one for capturing long-term user preferences and another for capturing short-term user preferences. During each update, the short-term LoRA module is temporarily introduced and trained on new data, while the long-term LoRA module remains fixed once trained on sufficient previous data. In the inference phase, the long-term LoRA module collaborates with the current short-term LoRA module to provide personalized recommendations. 
Next, we elaborate on the building of the two LoRA modules and the details of the inference:

\noindent \textbf{Short-term LoRA.} 
This LoRA module targets short-term preferences. Toward this, at each period $t$, we train a new module with parameters $\Theta_{t}$ using the newly collected data $\mathcal{D}_{t}$ as:
\begin{equation}
    \min_{\Theta_{t}} L (\mathcal{D}_{t} ;\Phi,\Theta_t ),
\end{equation}
where $\Phi$ represents the frozen pre-trained parameters of the LLM, and $L (\mathcal{D}_{t}; \Phi, \Theta_t )$ is the recommendation loss on $\mathcal{D}_{t}$. Notably, this new LoRA training approach is trained from scratch instead of being fine-tuned from the previous period, as fine-tuning has shown relatively poor performance in adapting to new preferences.

\noindent \textbf{Long-term LoRA.} 
This LoRA module aims to capture aggregated long-term preferences by fitting sufficient historical data. 
To achieve this, we train the long-term LoRA with ample historical data, denoted as $\mathcal{H} =\left \{\mathcal{D}_1, \mathcal{D}_2, \dots, \mathcal{D}_m \right \}$ as follows:
\begin{equation}
\label{eq:long-term-lora}
\min_{\Theta_{h}}L(\mathcal{H};\Phi, \Theta_{h}),
\vspace{-2pt}
\end{equation}
where $\Theta_{h}$ denotes long-term LoRA parameters. Once long-term LoRA is sufficiently trained after the $m$-th period, $\Theta_{h}$ can be updated at a slower pace, making LSAT training costs similar to fine-tuning. Before the $m$-th period, retraining $\Theta_{h}$ is needed.


\noindent \textbf{Inference.} During inference, we explore two methods to merge long- and short-term preferences from two LoRA modules:

\noindent \textit{1) Output ensemble:} 
This approach involves directly averaging the predictions with two LoRA modules. For a given sample $x$ at the $t$+1-th period, the final prediction is formulated as follows:
\begin{equation}
\label{eq:ensemble}
     \alpha f(x;\Phi, \Theta_{h})+ (1- \alpha)f(x;\Phi, \Theta_{t}),
\end{equation}
where $\alpha$ is a hyper-parameter chosen on the validation set, and $f(x;\Phi,\Theta_{t} )$ and $f(x;\Phi,\Theta_{h} )$ are predictions of LLM4Rec using the $t$-th period short-term LoRA and th long-term LoRA, respectively.

\noindent \textit{2) LoRA fusion:}
 As the output ensemble involves two LLM inferences, we explore merging the two LoRA modules for a single-pass inference, adopting a common fusion strategy---task arithmetic~\cite{zhang2023composing}.  
 Formally, task arithmetic fuses the parameters of the long-term LoRA ($\Theta_{h}$) and the short-term LoRA at the $t$-th period ($\Theta_{t}$), and generates the final prediction for a sample $x$ as:
\begin{equation}
\label{eq:ta}
    f(x;\Phi, \lambda \Theta_{h} + (1 - \lambda) \Theta_{t}),
\end{equation}
where $\lambda$ is a hyper-parameter chosen on the validation set.
\vspace{-6pt}
\section{Experiments}

\begin{figure}[t]
  \centering
    \includegraphics[width=0.45\textwidth]{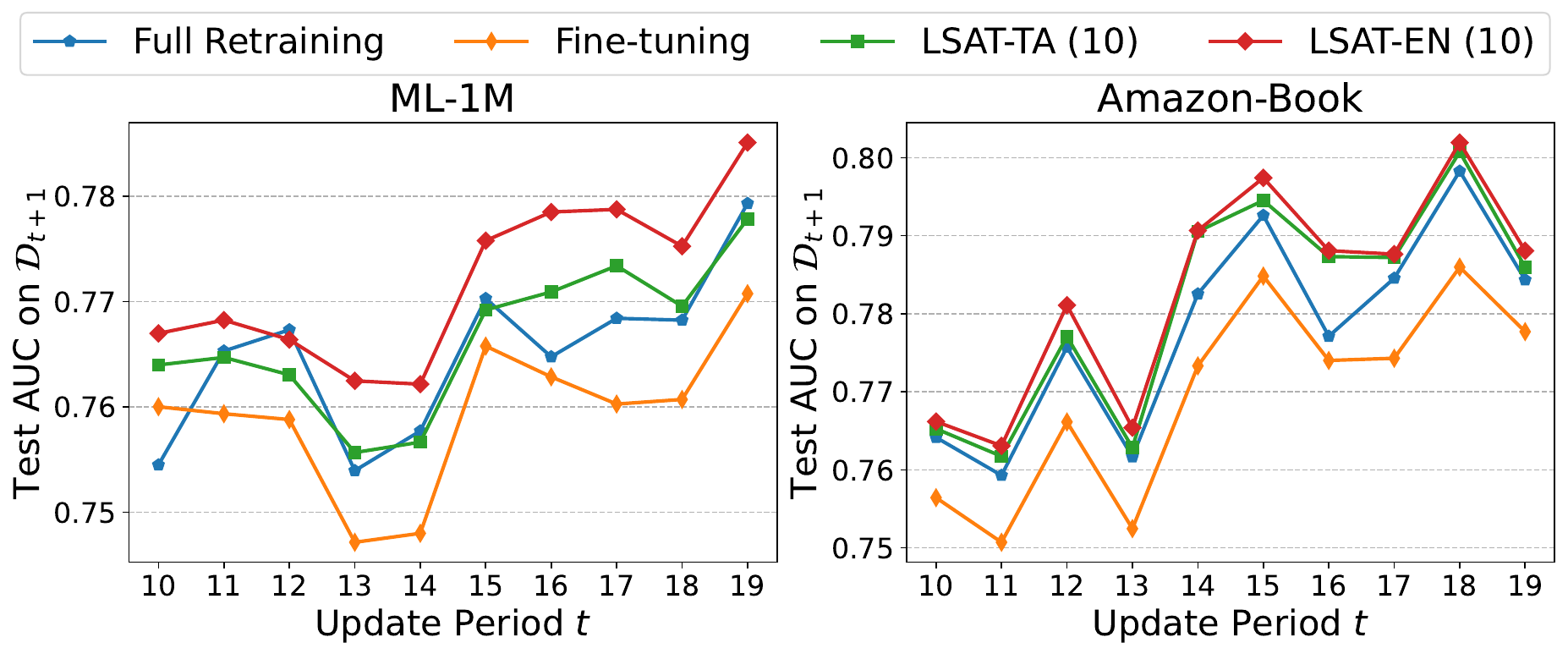}

    
  \vspace{-8pt}
  \caption{
  Performance comparison of full retraining, fine-tuning, and LSAT. 
  All models are updated promptly with newly collected data $\mathcal{D}_{t}$ and tested on $\mathcal{D}_{t+1}$. 
  LSAT (10) means it utilizes $\mathcal{H} =\left \{\mathcal{D}_1, \mathcal{D}_2, \dots, \mathcal{D}_{10} \right \}$ to train the long-term LoRA.
  }
  \vspace{-8pt}
  \label{fig:main-lsat}
\end{figure}

We conduct experiments to verify the effectiveness of our proposal.

\noindent \textbf{Experimental settings.}
We compare the performance of LSAT with full retraining and fine-tuning on ML-1M and Amazon-Book. 
Details about datasets and TALLRec are presented in Section~\ref{sec:setting1}.
To assess which approach yields superior
results after updates,
following~\cite{sml}, the model is updated with $\mathcal{D}_t$ and evaluated on $\mathcal{D}_{t+1}$.
For LSAT, the long-term LoRA module relies on a historical dataset $\mathcal{H} =\left \{\mathcal{D}_1, \mathcal{D}_2, \dots, \mathcal{D}_m \right \}$ (Equation~\eqref{eq:long-term-lora}), with $m$ set to 10 (LSAT (10)).
We study two methods for model merging defined in the Inference part of Section~\ref{sec:LSAT}: ensemble (LSAT-EN), and task arithmetic (LSAT-TA).
 The coefficient of LSAT $\alpha$ (Equation~\ref{eq:ensemble}) and $\lambda$ (Equation~\ref{eq:ta}) are searched within $\left \{0, 0.1, \dots , 1 \right \}$.

\noindent \textbf{Overall results.} 
Figure~\ref{fig:main-lsat} illustrates the overall performance comparison between full retraining, fine-tuning, and LSAT.
From the figure, we can find that LSAT-EN and LSAT-TA outperform full retraining and fine-tuning on two datasets, emphasizing the effectiveness of using separate LoRAs for long-term and short-term interests. This supports our initial hypothesis that employing two adapters leads to improved performance in modeling both long-term and short-term interests. Nevertheless, 
LSAT-TA's relatively modest improvement suggests merging adapter parameters into a single adapter may be less effective, implying the need for exploring a parameter-level LoRA merging method tailored for incremental learning in recommendation systems.
\begin{table}[t]
\centering
\caption{
Average AUC across $\mathcal{D}_{11}-\mathcal{D}_{20}$.
All models are updated promptly with newly collected data $\mathcal{D}_{t}$ and tested on $\mathcal{D}_{t+1}$. LSAT-EN (full) means its long-term LoRA is retrained during each period with all historical data.
}
\label{tab:ablation}
\vspace{-4pt}
\renewcommand\arraystretch{0.88}
	\setlength{\tabcolsep}{1mm}{
		\resizebox{0.4\textwidth}{!}{
  \small
\begin{tabular}{l|cc}
\hline
                & \multicolumn{1}{l}{ML-1M} & \multicolumn{1}{l}{Amazon-Book} \\ \hline
Full Retraining & 0.7650                    & 0.7780                          \\
Fine-tuning     & 0.7594                    & 0.7696                          \\
Short-term LoRA & 0.7638                    & 0.7806                          \\
Long-term LoRA  & 0.7617                    & 0.7789                          \\
LSAT-TA (10)  & 0.7665           
& 0.7813                           \\
LSAT-EN (full)  & 0.7691                    & 0.7822                          \\
LSAT-EN (10)    & \textbf{0.7720}           & \textbf{0.7830}                 \\
\hline
\end{tabular}
}}
\vspace{-6pt}
\end{table}

\noindent \textbf{Analyses.}
We then study LSAT's two integral components: the short-term LoRA and the long-term LoRA, comparing their performance with LSAT-EN in Table~\ref{tab:ablation}, where we can find: 
\begin{itemize}[leftmargin=*]
\vspace{-3pt}


\item
Short-term LoRA surpasses fine-tuning, as fine-tuning may suffer catastrophic forgetting~\cite{incrementalSurvey,luo2023empirical}, where newly acquired knowledge conflicts with old knowledge, resulting in performance decline.
This suggests the rationale of using a new LoRA to learn the latest preferences, rather than fine-tuning from the previous stage.
%

\item LSAT-EN (full), consistently updating the long-term LoRA, does not bring additional improvements compared to LSAT-EN (10), using a fixed long-term LoRA. These results suggest that the long-term LoRA can be updated at a slower pace once adequately trained, as long-term preferences tend to remain relatively stable.

\item Using only a single short-term LoRA or long-term LoRA leads to a performance decrease, highlighting the importance of merging both the long-term and short-term LoRAs for LSAT.

\vspace{-2pt}


\end{itemize}

\vspace{-9pt}
\section{Conclusion}

This study studies the impact of incremental learning on LLM4Rec. Empirical results and analysis reveal that both full retraining and fine-tuning fail to deliver the anticipated performance improvement for LLM4Rec. We posit that a singular LoRA may encounter challenges in simultaneously capturing long-term and short-term preferences. To validate our hypothesis, we introduce LSAT and conduct experimental validation. 
Nevertheless, our current research is limited to the TALLRec backbone, which uses solely textual information (title) for recommendations. Future investigations will extend to other backbones.
Additionally, as LSAT studies incremental learning methods solely from the perspective of LoRA capacity, we plan to explore other dimensions for more effective methods.



\begin{acks}
This work is supported by the National Key Research and Development Program of China (2022YFB3104701), the National Natural Science Foundation of China (62272437), and the CCCD Key Lab of Ministry of Culture and Tourism.
\end{acks}



\bibliographystyle{ACM-Reference-Format}

\bibliography{ref-bib} 


\begin{thebibliography}{45}


\ifx \showCODEN    \undefined \def \showCODEN     #1{\unskip}     \fi
\ifx \showDOI      \undefined \def \showDOI       #1{#1}\fi
\ifx \showISBNx    \undefined \def \showISBNx     #1{\unskip}     \fi
\ifx \showISBNxiii \undefined \def \showISBNxiii  #1{\unskip}     \fi
\ifx \showISSN     \undefined \def \showISSN      #1{\unskip}     \fi
\ifx \showLCCN     \undefined \def \showLCCN      #1{\unskip}     \fi
\ifx \shownote     \undefined \def \shownote      #1{#1}          \fi
\ifx \showarticletitle \undefined \def \showarticletitle #1{#1}   \fi
\ifx \showURL      \undefined \def \showURL       {\relax}        \fi
\providecommand\bibfield[2]{#2}
\providecommand\bibinfo[2]{#2}
\providecommand\natexlab[1]{#1}
\providecommand\showeprint[2][]{arXiv:#2}

\bibitem[Ai et~al\mbox{.}(2023)]%
        {LLMREC1}
\bibfield{author}{\bibinfo{person}{Qingyao Ai}, \bibinfo{person}{Ting Bai}, \bibinfo{person}{Zhao Cao}, \bibinfo{person}{Yi Chang}, \bibinfo{person}{Jiawei Chen}, \bibinfo{person}{Zhumin Chen}, \bibinfo{person}{Zhiyong Cheng}, \bibinfo{person}{Shoubin Dong}, \bibinfo{person}{Zhicheng Dou}, \bibinfo{person}{Fuli Feng}, {et~al\mbox{.}}} \bibinfo{year}{2023}\natexlab{}.
\newblock \showarticletitle{Information Retrieval Meets Large Language Models: A Strategic Report from Chinese IR Community}.
\newblock \bibinfo{journal}{\emph{AI Open}}  \bibinfo{volume}{4} (\bibinfo{year}{2023}), \bibinfo{pages}{80--90}.
\newblock


\bibitem[Bao et~al\mbox{.}(2023a)]%
        {bigrec}
\bibfield{author}{\bibinfo{person}{Keqin Bao}, \bibinfo{person}{Jizhi Zhang}, \bibinfo{person}{Wenjie Wang}, \bibinfo{person}{Yang Zhang}, \bibinfo{person}{Zhengyi Yang}, \bibinfo{person}{Yancheng Luo}, \bibinfo{person}{Fuli Feng}, \bibinfo{person}{Xiangnan He}, {and} \bibinfo{person}{Qi Tian}.} \bibinfo{year}{2023}\natexlab{a}.
\newblock \showarticletitle{A Bi-step Grounding Paradigm for Large Language Models in Recommendation Systems}.
\newblock \bibinfo{journal}{\emph{arXiv preprint arXiv:2308.08434}} (\bibinfo{year}{2023}).
\newblock


\bibitem[Bao et~al\mbox{.}(2023b)]%
        {tallrec}
\bibfield{author}{\bibinfo{person}{Keqin Bao}, \bibinfo{person}{Jizhi Zhang}, \bibinfo{person}{Yang Zhang}, \bibinfo{person}{Wenjie Wang}, \bibinfo{person}{Fuli Feng}, {and} \bibinfo{person}{Xiangnan He}.} \bibinfo{year}{2023}\natexlab{b}.
\newblock \showarticletitle{TALLRec: An Effective and Efficient Tuning Framework to Align Large Language Model with Recommendation}. In \bibinfo{booktitle}{\emph{Proceedings of the 17th ACM Conference on Recommender Systems}}. \bibinfo{pages}{1007–1014}.
\newblock


\bibitem[Berrios et~al\mbox{.}(2023)]%
        {berrios2023towards}
\bibfield{author}{\bibinfo{person}{William Berrios}, \bibinfo{person}{Gautam Mittal}, \bibinfo{person}{Tristan Thrush}, \bibinfo{person}{Douwe Kiela}, {and} \bibinfo{person}{Amanpreet Singh}.} \bibinfo{year}{2023}\natexlab{}.
\newblock \showarticletitle{Towards Language Models that can See: Computer Vision through the Lens of Natural Language}.
\newblock \bibinfo{journal}{\emph{arXiv preprint arXiv:2306.16410}} (\bibinfo{year}{2023}).
\newblock


\bibitem[Chen(2023)]%
        {chen2023palr}
\bibfield{author}{\bibinfo{person}{Zheng Chen}.} \bibinfo{year}{2023}\natexlab{}.
\newblock \showarticletitle{PALR: Personalization Aware LLMs for Recommendation}.
\newblock \bibinfo{journal}{\emph{arXiv preprint arXiv:2305.07622}} (\bibinfo{year}{2023}).
\newblock


\bibitem[Chronopoulou et~al\mbox{.}(2023)]%
        {chronopoulou2023adaptersoup}
\bibfield{author}{\bibinfo{person}{Alexandra Chronopoulou}, \bibinfo{person}{Matthew~E Peters}, \bibinfo{person}{Alexander Fraser}, {and} \bibinfo{person}{Jesse Dodge}.} \bibinfo{year}{2023}\natexlab{}.
\newblock \showarticletitle{Adaptersoup: Weight Averaging to Improve Generalization of Pretrained Language Models}.
\newblock \bibinfo{journal}{\emph{arXiv preprint arXiv:2302.07027}} (\bibinfo{year}{2023}).
\newblock


\bibitem[Diaz-Aviles et~al\mbox{.}(2012)]%
        {diaz2012real}
\bibfield{author}{\bibinfo{person}{Ernesto Diaz-Aviles}, \bibinfo{person}{Lucas Drumond}, \bibinfo{person}{Lars Schmidt-Thieme}, {and} \bibinfo{person}{Wolfgang Nejdl}.} \bibinfo{year}{2012}\natexlab{}.
\newblock \showarticletitle{Real-time Top-n Recommendation in Social Streams}. In \bibinfo{booktitle}{\emph{Proceedings of the sixth ACM Conference on Recommender Systems}}. \bibinfo{pages}{59--66}.
\newblock


\bibitem[Ding et~al\mbox{.}(2023)]%
        {liu-peft}
\bibfield{author}{\bibinfo{person}{Ning Ding}, \bibinfo{person}{Yujia Qin}, \bibinfo{person}{Guang Yang}, \bibinfo{person}{Fuchao Wei}, \bibinfo{person}{Zonghan Yang}, \bibinfo{person}{Yusheng Su}, \bibinfo{person}{Shengding Hu}, \bibinfo{person}{Yulin Chen}, \bibinfo{person}{Chi-Min Chan}, \bibinfo{person}{Weize Chen}, {et~al\mbox{.}}} \bibinfo{year}{2023}\natexlab{}.
\newblock \showarticletitle{Parameter-efficient Fine-tuning of Large-scale Pre-trained Language Models}.
\newblock \bibinfo{journal}{\emph{Nature Machine Intelligence}} \bibinfo{volume}{5}, \bibinfo{number}{3} (\bibinfo{year}{2023}), \bibinfo{pages}{220--235}.
\newblock


\bibitem[Driess et~al\mbox{.}(2023)]%
        {palm}
\bibfield{author}{\bibinfo{person}{Danny Driess}, \bibinfo{person}{Fei Xia}, \bibinfo{person}{Mehdi~SM Sajjadi}, \bibinfo{person}{Corey Lynch}, \bibinfo{person}{Aakanksha Chowdhery}, \bibinfo{person}{Brian Ichter}, \bibinfo{person}{Ayzaan Wahid}, \bibinfo{person}{Jonathan Tompson}, \bibinfo{person}{Quan Vuong}, \bibinfo{person}{Tianhe Yu}, {et~al\mbox{.}}} \bibinfo{year}{2023}\natexlab{}.
\newblock \showarticletitle{Palm-e: An Embodied Multimodal Language Model}.
\newblock \bibinfo{journal}{\emph{arXiv preprint arXiv:2303.03378}} (\bibinfo{year}{2023}).
\newblock


\bibitem[Fan et~al\mbox{.}(2023)]%
        {llmrec2}
\bibfield{author}{\bibinfo{person}{Wenqi Fan}, \bibinfo{person}{Zihuai Zhao}, \bibinfo{person}{Jiatong Li}, \bibinfo{person}{Yunqing Liu}, \bibinfo{person}{Xiaowei Mei}, \bibinfo{person}{Yiqi Wang}, \bibinfo{person}{Jiliang Tang}, {and} \bibinfo{person}{Qing Li}.} \bibinfo{year}{2023}\natexlab{}.
\newblock \showarticletitle{Recommender Systems in the Era of Large Language Models}.
\newblock \bibinfo{journal}{\emph{arXiv preprint arXiv:2307.02046}} (\bibinfo{year}{2023}).
\newblock


\bibitem[Hadi et~al\mbox{.}(2023)]%
        {surveyLLM}
\bibfield{author}{\bibinfo{person}{Muhammad~Usman Hadi}, \bibinfo{person}{R Qureshi}, \bibinfo{person}{A Shah}, \bibinfo{person}{M Irfan}, \bibinfo{person}{A Zafar}, \bibinfo{person}{MB Shaikh}, \bibinfo{person}{N Akhtar}, \bibinfo{person}{J Wu}, {and} \bibinfo{person}{S Mirjalili}.} \bibinfo{year}{2023}\natexlab{}.
\newblock \showarticletitle{A Survey on Large Language Models: Applications, Challenges, Limitations, and Practical Usage}.
\newblock \bibinfo{journal}{\emph{TechRxiv}} (\bibinfo{year}{2023}).
\newblock


\bibitem[Hanley and McNeil(1982)]%
        {auc}
\bibfield{author}{\bibinfo{person}{James~A Hanley} {and} \bibinfo{person}{Barbara~J McNeil}.} \bibinfo{year}{1982}\natexlab{}.
\newblock \showarticletitle{The Meaning and Use of the Area under a Receiver Operating Characteristic (ROC) Curve.}
\newblock \bibinfo{journal}{\emph{Radiology}} \bibinfo{volume}{143}, \bibinfo{number}{1} (\bibinfo{year}{1982}), \bibinfo{pages}{29--36}.
\newblock


\bibitem[Harper and Konstan(2015)]%
        {movielens}
\bibfield{author}{\bibinfo{person}{F~Maxwell Harper} {and} \bibinfo{person}{Joseph~A Konstan}.} \bibinfo{year}{2015}\natexlab{}.
\newblock \showarticletitle{The Movielens Datasets: History and Context}.
\newblock \bibinfo{journal}{\emph{Acm Transactions on Interactive Intelligent Systems}} \bibinfo{volume}{5}, \bibinfo{number}{4} (\bibinfo{year}{2015}), \bibinfo{pages}{1--19}.
\newblock


\bibitem[Hidasi et~al\mbox{.}(2016)]%
        {gru4rec}
\bibfield{author}{\bibinfo{person}{Balazs Hidasi}, \bibinfo{person}{Alexandros Karatzoglou}, \bibinfo{person}{Linas Baltrunas}, {and} \bibinfo{person}{Domonkos Tikk}.} \bibinfo{year}{2016}\natexlab{}.
\newblock \showarticletitle{Session-based Recommendations with Recurrent Neural Networks}. In \bibinfo{booktitle}{\emph{4th International Conference on Learning Representations, {ICLR} 2016}}.
\newblock


\bibitem[Hu et~al\mbox{.}(2021)]%
        {lora}
\bibfield{author}{\bibinfo{person}{Edward~J Hu}, \bibinfo{person}{Phillip Wallis}, \bibinfo{person}{Zeyuan Allen-Zhu}, \bibinfo{person}{Yuanzhi Li}, \bibinfo{person}{Shean Wang}, \bibinfo{person}{Lu Wang}, \bibinfo{person}{Weizhu Chen}, {et~al\mbox{.}}} \bibinfo{year}{2021}\natexlab{}.
\newblock \showarticletitle{LoRA: Low-Rank Adaptation of Large Language Models}. In \bibinfo{booktitle}{\emph{International Conference on Learning Representations}}.
\newblock


\bibitem[Jang et~al\mbox{.}(2021)]%
        {jang2021towards}
\bibfield{author}{\bibinfo{person}{Joel Jang}, \bibinfo{person}{Seonghyeon Ye}, \bibinfo{person}{Sohee Yang}, \bibinfo{person}{Joongbo Shin}, \bibinfo{person}{Janghoon Han}, \bibinfo{person}{KIM Gyeonghun}, \bibinfo{person}{Stanley~Jungkyu Choi}, {and} \bibinfo{person}{Minjoon Seo}.} \bibinfo{year}{2021}\natexlab{}.
\newblock \showarticletitle{Towards Continual Knowledge Learning of Language Models}. In \bibinfo{booktitle}{\emph{International Conference on Learning Representations}}.
\newblock


\bibitem[Kang and McAuley(2018)]%
        {kang2018self}
\bibfield{author}{\bibinfo{person}{Wang-Cheng Kang} {and} \bibinfo{person}{Julian McAuley}.} \bibinfo{year}{2018}\natexlab{}.
\newblock \showarticletitle{Self-attentive Sequential Recommendation}. In \bibinfo{booktitle}{\emph{2018 IEEE International Conference on Data Mining (ICDM)}}. IEEE, \bibinfo{pages}{197--206}.
\newblock


\bibitem[Kingma and Ba(2014)]%
        {kingma2014adam}
\bibfield{author}{\bibinfo{person}{Diederik~P Kingma} {and} \bibinfo{person}{Jimmy Ba}.} \bibinfo{year}{2014}\natexlab{}.
\newblock \showarticletitle{Adam: A Method for Stochastic Optimization}.
\newblock \bibinfo{journal}{\emph{arXiv preprint arXiv:1412.6980}} (\bibinfo{year}{2014}).
\newblock


\bibitem[Koren et~al\mbox{.}(2009)]%
        {koren2009matrix}
\bibfield{author}{\bibinfo{person}{Yehuda Koren}, \bibinfo{person}{Robert Bell}, {and} \bibinfo{person}{Chris Volinsky}.} \bibinfo{year}{2009}\natexlab{}.
\newblock \showarticletitle{Matrix Factorization Techniques for Recommender Systems}.
\newblock \bibinfo{journal}{\emph{Computer}} \bibinfo{volume}{42}, \bibinfo{number}{8} (\bibinfo{year}{2009}), \bibinfo{pages}{30--37}.
\newblock


\bibitem[Lee et~al\mbox{.}(2023)]%
        {period-update}
\bibfield{author}{\bibinfo{person}{Hyunsung Lee}, \bibinfo{person}{Sungwook Yoo}, \bibinfo{person}{Dongjun Lee}, {and} \bibinfo{person}{Jaekwang Kim}.} \bibinfo{year}{2023}\natexlab{}.
\newblock \showarticletitle{How Important is Periodic Model Update in Recommender System?}. In \bibinfo{booktitle}{\emph{Proceedings of the 46th International ACM SIGIR Conference on Research and Development in Information Retrieval}}. \bibinfo{pages}{2661--2668}.
\newblock


\bibitem[Li et~al\mbox{.}(2023)]%
        {bookgpt}
\bibfield{author}{\bibinfo{person}{Zhiyu Li}, \bibinfo{person}{Yanfang Chen}, \bibinfo{person}{Xuan Zhang}, {and} \bibinfo{person}{Xun Liang}.} \bibinfo{year}{2023}\natexlab{}.
\newblock \showarticletitle{BookGPT: A General Framework for Book Recommendation Empowered by Large Language Model}.
\newblock \bibinfo{journal}{\emph{Electronics}} \bibinfo{volume}{12}, \bibinfo{number}{22} (\bibinfo{year}{2023}), \bibinfo{pages}{4654}.
\newblock


\bibitem[Lin et~al\mbox{.}(2023a)]%
        {lin2023can}
\bibfield{author}{\bibinfo{person}{Jianghao Lin}, \bibinfo{person}{Xinyi Dai}, \bibinfo{person}{Yunjia Xi}, \bibinfo{person}{Weiwen Liu}, \bibinfo{person}{Bo Chen}, \bibinfo{person}{Xiangyang Li}, \bibinfo{person}{Chenxu Zhu}, \bibinfo{person}{Huifeng Guo}, \bibinfo{person}{Yong Yu}, \bibinfo{person}{Ruiming Tang}, {et~al\mbox{.}}} \bibinfo{year}{2023}\natexlab{a}.
\newblock \showarticletitle{How Can Recommender Systems Benefit from Large Language Models: A Survey}.
\newblock \bibinfo{journal}{\emph{arXiv preprint arXiv:2306.05817}} (\bibinfo{year}{2023}).
\newblock


\bibitem[Lin et~al\mbox{.}(2023b)]%
        {rella}
\bibfield{author}{\bibinfo{person}{Jianghao Lin}, \bibinfo{person}{Rong Shan}, \bibinfo{person}{Chenxu Zhu}, \bibinfo{person}{Kounianhua Du}, \bibinfo{person}{Bo Chen}, \bibinfo{person}{Shigang Quan}, \bibinfo{person}{Ruiming Tang}, \bibinfo{person}{Yong Yu}, {and} \bibinfo{person}{Weinan Zhang}.} \bibinfo{year}{2023}\natexlab{b}.
\newblock \showarticletitle{ReLLa: Retrieval-enhanced Large Language Models for Lifelong Sequential Behavior Comprehension in Recommendation}.
\newblock \bibinfo{journal}{\emph{arXiv preprint arXiv:2308.11131}} (\bibinfo{year}{2023}).
\newblock


\bibitem[Liu et~al\mbox{.}(2023)]%
        {llmrec-benchmark}
\bibfield{author}{\bibinfo{person}{Junling Liu}, \bibinfo{person}{Chao Liu}, \bibinfo{person}{Peilin Zhou}, \bibinfo{person}{Qichen Ye}, \bibinfo{person}{Dading Chong}, \bibinfo{person}{Kang Zhou}, \bibinfo{person}{Yueqi Xie}, \bibinfo{person}{Yuwei Cao}, \bibinfo{person}{Shoujin Wang}, \bibinfo{person}{Chenyu You}, {et~al\mbox{.}}} \bibinfo{year}{2023}\natexlab{}.
\newblock \showarticletitle{LLMRec: Benchmarking Large Language Models on Recommendation Task}.
\newblock \bibinfo{journal}{\emph{arXiv preprint arXiv:2308.12241}} (\bibinfo{year}{2023}).
\newblock


\bibitem[Luo et~al\mbox{.}(2023)]%
        {luo2023empirical}
\bibfield{author}{\bibinfo{person}{Yun Luo}, \bibinfo{person}{Zhen Yang}, \bibinfo{person}{Fandong Meng}, \bibinfo{person}{Yafu Li}, \bibinfo{person}{Jie Zhou}, {and} \bibinfo{person}{Yue Zhang}.} \bibinfo{year}{2023}\natexlab{}.
\newblock \showarticletitle{An Empirical Study of Catastrophic Forgetting in Large Language Models during Continual Fine-tuning}.
\newblock \bibinfo{journal}{\emph{arXiv preprint arXiv:2308.08747}} (\bibinfo{year}{2023}).
\newblock


\bibitem[Ni et~al\mbox{.}(2019)]%
        {amazonbook}
\bibfield{author}{\bibinfo{person}{Jianmo Ni}, \bibinfo{person}{Jiacheng Li}, {and} \bibinfo{person}{Julian McAuley}.} \bibinfo{year}{2019}\natexlab{}.
\newblock \showarticletitle{Justifying Recommendations Using Distantly-labeled Reviews and Fine-grained Aspects}. In \bibinfo{booktitle}{\emph{Proceedings of the 2019 Conference on Empirical Methods in Natural Language Processing and the 9th International Joint Conference on Natural Language Processing (EMNLP-IJCNLP)}}. \bibinfo{pages}{188--197}.
\newblock


\bibitem[Ouyang et~al\mbox{.}(2022)]%
        {instruction-tuning}
\bibfield{author}{\bibinfo{person}{Long Ouyang}, \bibinfo{person}{Jeffrey Wu}, \bibinfo{person}{Xu Jiang}, \bibinfo{person}{Diogo Almeida}, \bibinfo{person}{Carroll Wainwright}, \bibinfo{person}{Pamela Mishkin}, \bibinfo{person}{Chong Zhang}, \bibinfo{person}{Sandhini Agarwal}, \bibinfo{person}{Katarina Slama}, \bibinfo{person}{Alex Ray}, {et~al\mbox{.}}} \bibinfo{year}{2022}\natexlab{}.
\newblock \showarticletitle{Training Language Models to Follow Instructions with Human Feedback}.
\newblock \bibinfo{journal}{\emph{Advances in Neural Information Processing Systems}}  \bibinfo{volume}{35} (\bibinfo{year}{2022}), \bibinfo{pages}{27730--27744}.
\newblock


\bibitem[Rendle and Schmidt-Thieme(2008)]%
        {rendle2008online}
\bibfield{author}{\bibinfo{person}{Steffen Rendle} {and} \bibinfo{person}{Lars Schmidt-Thieme}.} \bibinfo{year}{2008}\natexlab{}.
\newblock \showarticletitle{Online-updating Regularized Kernel Matrix Factorization Models for Large-scale Recommender systems}. In \bibinfo{booktitle}{\emph{Proceedings of the 2008 ACM conference on Recommender systems}}. \bibinfo{pages}{251--258}.
\newblock


\bibitem[Sima et~al\mbox{.}(2022)]%
        {sima2022ekko}
\bibfield{author}{\bibinfo{person}{Chijun Sima}, \bibinfo{person}{Yao Fu}, \bibinfo{person}{Man-Kit Sit}, \bibinfo{person}{Liyi Guo}, \bibinfo{person}{Xuri Gong}, \bibinfo{person}{Feng Lin}, \bibinfo{person}{Junyu Wu}, \bibinfo{person}{Yongsheng Li}, \bibinfo{person}{Haidong Rong}, \bibinfo{person}{Pierre-Louis Aublin}, {et~al\mbox{.}}} \bibinfo{year}{2022}\natexlab{}.
\newblock \showarticletitle{Ekko: A $\{$Large-Scale$\}$ Deep Learning Recommender System with $\{$Low-Latency$\}$ Model Update}. In \bibinfo{booktitle}{\emph{16th USENIX Symposium on Operating Systems Design and Implementation (OSDI 22)}}. \bibinfo{pages}{821--839}.
\newblock


\bibitem[Singh et~al\mbox{.}(2023)]%
        {singh2023progprompt}
\bibfield{author}{\bibinfo{person}{Ishika Singh}, \bibinfo{person}{Valts Blukis}, \bibinfo{person}{Arsalan Mousavian}, \bibinfo{person}{Ankit Goyal}, \bibinfo{person}{Danfei Xu}, \bibinfo{person}{Jonathan Tremblay}, \bibinfo{person}{Dieter Fox}, \bibinfo{person}{Jesse Thomason}, {and} \bibinfo{person}{Animesh Garg}.} \bibinfo{year}{2023}\natexlab{}.
\newblock \showarticletitle{Progprompt: Generating Situated Robot Task Plans Using Large Language Models}. In \bibinfo{booktitle}{\emph{2023 IEEE International Conference on Robotics and Automation (ICRA)}}. IEEE, \bibinfo{pages}{11523--11530}.
\newblock


\bibitem[Tang and Wang(2018)]%
        {tang2018personalized}
\bibfield{author}{\bibinfo{person}{Jiaxi Tang} {and} \bibinfo{person}{Ke Wang}.} \bibinfo{year}{2018}\natexlab{}.
\newblock \showarticletitle{Personalized Top-n Sequential Recommendation via Convolutional Sequence Embedding}. In \bibinfo{booktitle}{\emph{Proceedings of the Eleventh ACM International Conference on Web Search and Data Mining}}. \bibinfo{pages}{565--573}.
\newblock


\bibitem[Touvron et~al\mbox{.}(2023)]%
        {touvron2023llama}
\bibfield{author}{\bibinfo{person}{Hugo Touvron}, \bibinfo{person}{Thibaut Lavril}, \bibinfo{person}{Gautier Izacard}, \bibinfo{person}{Xavier Martinet}, \bibinfo{person}{Marie-Anne Lachaux}, \bibinfo{person}{Timoth{\'e}e Lacroix}, \bibinfo{person}{Baptiste Rozi{\`e}re}, \bibinfo{person}{Naman Goyal}, \bibinfo{person}{Eric Hambro}, \bibinfo{person}{Faisal Azhar}, {et~al\mbox{.}}} \bibinfo{year}{2023}\natexlab{}.
\newblock \showarticletitle{LLaMA: Open and Efficient Foundation Language Models}.
\newblock \bibinfo{journal}{\emph{arXiv preprint arXiv:2302.13971}} (\bibinfo{year}{2023}).
\newblock


\bibitem[Van~den Oord et~al\mbox{.}(2013)]%
        {deepmusic}
\bibfield{author}{\bibinfo{person}{Aaron Van~den Oord}, \bibinfo{person}{Sander Dieleman}, {and} \bibinfo{person}{Benjamin Schrauwen}.} \bibinfo{year}{2013}\natexlab{}.
\newblock \showarticletitle{Deep Content-based Music Recommendation}.
\newblock \bibinfo{journal}{\emph{Advances in Neural Information Processing Systems}}  \bibinfo{volume}{26} (\bibinfo{year}{2013}).
\newblock


\bibitem[Wang and Lim(2023)]%
        {wang2023zero}
\bibfield{author}{\bibinfo{person}{Lei Wang} {and} \bibinfo{person}{Ee-Peng Lim}.} \bibinfo{year}{2023}\natexlab{}.
\newblock \showarticletitle{Zero-Shot Next-Item Recommendation using Large Pretrained Language Models}.
\newblock \bibinfo{journal}{\emph{arXiv preprint arXiv:2304.03153}} (\bibinfo{year}{2023}).
\newblock


\bibitem[Wang et~al\mbox{.}(2018a)]%
        {wang2018neural}
\bibfield{author}{\bibinfo{person}{Qinyong Wang}, \bibinfo{person}{Hongzhi Yin}, \bibinfo{person}{Zhiting Hu}, \bibinfo{person}{Defu Lian}, \bibinfo{person}{Hao Wang}, {and} \bibinfo{person}{Zi Huang}.} \bibinfo{year}{2018}\natexlab{a}.
\newblock \showarticletitle{Neural Memory Streaming Recommender Networks with Adversarial Training}. In \bibinfo{booktitle}{\emph{Proceedings of the 24th ACM SIGKDD International Conference on Knowledge Discovery \& Data Mining}}. \bibinfo{pages}{2467--2475}.
\newblock


\bibitem[Wang et~al\mbox{.}(2018b)]%
        {wang2018streaming}
\bibfield{author}{\bibinfo{person}{Weiqing Wang}, \bibinfo{person}{Hongzhi Yin}, \bibinfo{person}{Zi Huang}, \bibinfo{person}{Qinyong Wang}, \bibinfo{person}{Xingzhong Du}, {and} \bibinfo{person}{Quoc Viet~Hung Nguyen}.} \bibinfo{year}{2018}\natexlab{b}.
\newblock \showarticletitle{Streaming Ranking Based Recommender Systems}. In \bibinfo{booktitle}{\emph{The 41st International ACM SIGIR Conference on Research \& Development in Information Retrieval}}. \bibinfo{pages}{525--534}.
\newblock


\bibitem[Wu et~al\mbox{.}(2023)]%
        {llmrec3}
\bibfield{author}{\bibinfo{person}{Likang Wu}, \bibinfo{person}{Zhi Zheng}, \bibinfo{person}{Zhaopeng Qiu}, \bibinfo{person}{Hao Wang}, \bibinfo{person}{Hongchao Gu}, \bibinfo{person}{Tingjia Shen}, \bibinfo{person}{Chuan Qin}, \bibinfo{person}{Chen Zhu}, \bibinfo{person}{Hengshu Zhu}, \bibinfo{person}{Qi Liu}, {et~al\mbox{.}}} \bibinfo{year}{2023}\natexlab{}.
\newblock \showarticletitle{A Survey on Large Language Models for Recommendation}.
\newblock \bibinfo{journal}{\emph{arXiv preprint arXiv:2305.19860}} (\bibinfo{year}{2023}).
\newblock


\bibitem[Xie et~al\mbox{.}(2023)]%
        {cvllm}
\bibfield{author}{\bibinfo{person}{Lingxi Xie}, \bibinfo{person}{Longhui Wei}, \bibinfo{person}{Xiaopeng Zhang}, \bibinfo{person}{Kaifeng Bi}, \bibinfo{person}{Xiaotao Gu}, \bibinfo{person}{Jianlong Chang}, {and} \bibinfo{person}{Qi Tian}.} \bibinfo{year}{2023}\natexlab{}.
\newblock \showarticletitle{Towards AGI in Computer Vision: Lessons Learned from GPT and Large Language Models}.
\newblock \bibinfo{journal}{\emph{arXiv preprint arXiv:2306.08641}} (\bibinfo{year}{2023}).
\newblock


\bibitem[Xie et~al\mbox{.}(2022)]%
        {longshort}
\bibfield{author}{\bibinfo{person}{Ruobing Xie}, \bibinfo{person}{Yalong Wang}, \bibinfo{person}{Rui Wang}, \bibinfo{person}{Yuanfu Lu}, \bibinfo{person}{Yuanhang Zou}, \bibinfo{person}{Feng Xia}, {and} \bibinfo{person}{Leyu Lin}.} \bibinfo{year}{2022}\natexlab{}.
\newblock \showarticletitle{Long Short-term Temporal Meta-learning in Online Recommendation}. In \bibinfo{booktitle}{\emph{Proceedings of the Fifteenth ACM International Conference on Web Search and Data Mining}}. \bibinfo{pages}{1168--1176}.
\newblock


\bibitem[Zhang et~al\mbox{.}(2023a)]%
        {zhang2023composing}
\bibfield{author}{\bibinfo{person}{Jinghan Zhang}, \bibinfo{person}{Shiqi Chen}, \bibinfo{person}{Junteng Liu}, {and} \bibinfo{person}{Junxian He}.} \bibinfo{year}{2023}\natexlab{a}.
\newblock \showarticletitle{Composing Parameter-efficient Modules with Arithmetic Operations}.
\newblock \bibinfo{journal}{\emph{arXiv preprint arXiv:2306.14870}} (\bibinfo{year}{2023}).
\newblock


\bibitem[Zhang et~al\mbox{.}(2023c)]%
        {zhang2023recommendation}
\bibfield{author}{\bibinfo{person}{Junjie Zhang}, \bibinfo{person}{Ruobing Xie}, \bibinfo{person}{Yupeng Hou}, \bibinfo{person}{Wayne~Xin Zhao}, \bibinfo{person}{Leyu Lin}, {and} \bibinfo{person}{Ji-Rong Wen}.} \bibinfo{year}{2023}\natexlab{c}.
\newblock \showarticletitle{Recommendation as Instruction Following: A Large Language Model Empowered Recommendation Approach}.
\newblock \bibinfo{journal}{\emph{arXiv preprint arXiv:2305.07001}} (\bibinfo{year}{2023}).
\newblock


\bibitem[Zhang and Kim(2023)]%
        {incrementalSurvey}
\bibfield{author}{\bibinfo{person}{Peiyan Zhang} {and} \bibinfo{person}{Sunghun Kim}.} \bibinfo{year}{2023}\natexlab{}.
\newblock \showarticletitle{A Survey on Incremental Update for Neural Recommender Systems}.
\newblock \bibinfo{journal}{\emph{arXiv preprint arXiv:2303.02851}} (\bibinfo{year}{2023}).
\newblock


\bibitem[Zhang et~al\mbox{.}(2024)]%
        {zhang2024text}
\bibfield{author}{\bibinfo{person}{Yang Zhang}, \bibinfo{person}{Keqin Bao}, \bibinfo{person}{Ming Yan}, \bibinfo{person}{Wenjie Wang}, \bibinfo{person}{Fuli Feng}, {and} \bibinfo{person}{Xiangnan He}.} \bibinfo{year}{2024}\natexlab{}.
\newblock \showarticletitle{Text-like Encoding of Collaborative Information in Large Language Models for Recommendation}.
\newblock \bibinfo{journal}{\emph{arXiv preprint arXiv:2406.03210}} (\bibinfo{year}{2024}).
\newblock


\bibitem[Zhang et~al\mbox{.}(2020)]%
        {sml}
\bibfield{author}{\bibinfo{person}{Yang Zhang}, \bibinfo{person}{Fuli Feng}, \bibinfo{person}{Chenxu Wang}, \bibinfo{person}{Xiangnan He}, \bibinfo{person}{Meng Wang}, \bibinfo{person}{Yan Li}, {and} \bibinfo{person}{Yongdong Zhang}.} \bibinfo{year}{2020}\natexlab{}.
\newblock \showarticletitle{How to Retrain Recommender System? A Sequential Meta-learning Method}. In \bibinfo{booktitle}{\emph{Proceedings of the 43rd International ACM SIGIR Conference on Research and Development in Information Retrieval}}. \bibinfo{pages}{1479--1488}.
\newblock


\bibitem[Zhang et~al\mbox{.}(2023b)]%
        {collm}
\bibfield{author}{\bibinfo{person}{Yang Zhang}, \bibinfo{person}{Fuli Feng}, \bibinfo{person}{Jizhi Zhang}, \bibinfo{person}{Keqin Bao}, \bibinfo{person}{Qifan Wang}, {and} \bibinfo{person}{Xiangnan He}.} \bibinfo{year}{2023}\natexlab{b}.
\newblock \showarticletitle{CoLLM: Integrating Collaborative Embeddings into Large Language Models for Recommendation}.
\newblock \bibinfo{journal}{\emph{arXiv preprint arXiv:2310.19488}} (\bibinfo{year}{2023}).
\newblock


\end{thebibliography}
\end{document}